\documentclass[useAMS,usenatbib]{mn2e}
\usepackage{epsfig,lscape} 
\usepackage[dvipdfm]{hyperref} 
\usepackage{natbib}
\usepackage{booktabs}
\newcommand\mc[1]{\multicolumn{1}{c}{#1}}

\newcommand{\feii}{\hbox{[Fe\,{\sc ii}]}}

\title[Surface brightness profiles of BCDs]
{Surface brightness profiles of blue compact dwarf galaxies in the GOODS-N and GOODS-S field}

\author[J. H. Lian et al.]
{J. H. Lian\thanks{E-mail:ljhhw@mail.ustc.edu.cn (JHL);
xkong@ustc.edu.cn (XK)},
X.Kong, N.Jiang, W. Yan and Y. L. Gao\\
Department of Astronomy, University of Science and Technology of China, Hefei 230026, China\\
Key Laboratory for Research in Galaxies and Cosmology, Chinese Academy of Sciences, Hefei 230026, China}

\begin{document}

\maketitle

\begin{abstract}
We investigate the structural properties of the underlying hosts of 34 blue compact dwarf (BCD) galaxies with deep near-infrared (NIR) photometry. 
The BCD sample is selected from 
the Cosmic Assembly Near-IR Deep Extragalactic Legacy Survey in 
the Great observatories origins Deep Survey North and South
fields. We extract the surface
brightness profile (SBP) in the optical $F435W$ and NIR $F160W$ bands. 
The SBPs of BCDs in the $H$ band reach
$\sim$ 26 mag ${\rm arcsec^{-2}}$ at the 3$\sigma$ level, which is so far the deepest NIR imaging of BCDs.
Then we fit the SBPs with one- and two-component S\'ersic models.
About half of the BCDs favour the two-component model which significantly improves the fit quality.
The effective radii of the underlying hosts of BCDs in the $B$ band are smaller than those of
early-type dwarfs (dEs) and dwarf irregulars at a fixed luminosity.
This discrepancy is similar to findings in many previous works.
However,  
the difference in structural parameters between BCDs and other dwarf galaxies
seems to be less significant in the $H$ band.
Furthermore, we find a remarkable agreement between the underlying hosts of BCDs and dEs.
All dwarf galaxies seem 
to follow a similar luminosity-radius relationship which suggests a unified structural evolution for dwarf galaxies.
We conclude that a possible evolution track from BCDs to dEs cannot be ruled out, with no significant 
change of structure needed in the evolutionary scenario.
\end{abstract}

\begin{keywords}
galaxies: dwarf -- galaxies: evolution -- galaxies: photometry -- galaxies: structure
\end{keywords}

\section{introduction}

Blue compact dwarf (BCD) galaxies are compact and low-luminosity ($M_{\rm B}\ge-17$ mag) objects, which {are undergoing} active star 
formation at {the} present time \citep{sergent1970,kunth2000}. They are believed to be part of the building blocks of massive galaxies in {the} early 
Universe. The gas {consumption} time-scale of BCDs, {given the} current star formation rate, is much shorter than the age of the Universe \citep{kong2004}. 
BCDs {and dwarf irregulars (dIrrs)} are chemically unevolved and metal-poor ($Z_{\odot}/50\le Z\le Z_{\odot}/3$) galaxies 
{, early-type dwarfs (dEs)
have a considerable spread in metallicity \citep{kunth2000}}. 
{Some extremely metal-poor} BCDs were initially considered to be young galaxies 
with their first generation of stars forming. However, subsequent detection of the underlying extended old stellar population in most BCDs indicates that 
they are actually old systems \citep{papaderos1996,cairo2003,kong2003}. 
{The underlying old populations of BCDs are often regarded as the `underlying host' of the central star-forming region.}

The {evolutionary relationship} between BCDs and other dwarf galaxies is not clear and still under debate. 
A possible evolution scenario, proposed by \citet{thuan1985} and \citet{davies1988}, {is}
that {dIrrs} evolve into BCDs in several starbursts, enrich the interstellar, 
exhaust {their} gas and then fade to gas-free {dEs}.
Later on,
\citet{papaderos1996} discussed
the possible {evolutionary} connections between BCDs and other dwarf galaxies and suggested {that} the stellar wind from the central starburst is 
{necessary} to
explain the {compactness and central brightness} of BCDs. 
Recently, \citet{janowiecki2014} compared the structural properties of the underlying {hosts} of BCDs {in the} optical and near-infrared {(NIR)} bands
with {those} of dIrrs. They fit the outer {regions} of BCDs {exponentially} and
found that the 
underlying {hosts} of BCDs are brighter in the {centre} and have {smaller radii} than dIrrs. 
They concluded that, if BCDs can evolve into dIrrs, some physical 
mechanism should be responsible for the significant change of structural properties from BCDs to dIrrs. 
A {different} result was found by \citet{Micheva2013} with deep imaging of {luminous dwarf galaxies and intermediate-mass BCGs}.
They measured the structural parameters of the underlying {hosts} of BCGs with surface brightness {in the $B$} band $\mu_{\rm B}$ in {the range} 
26--28~${\rm mag\ arcsec^{-2}}$
and found {that} the hosts of BCGs {have similar structural parameters}
with other dwarf galaxies like dEs and dIrrs. 
They suggested that the 
{obtained} structural 
properties {for} surface brightness $\mu_{\rm B}$ in 24--26~${\rm mag\ arcsec^{-2}}$ {range} are {possibly} affected by nebular emission
{from central star-forming regions}. 
It should be 
noted that the depth of the {$B$} band image in \citet{janowiecki2014} only reaches $\sim$ 26~mag ${\rm arcsec^{-2}}$. 
Meanwhile, a fit only to the outer region 
of {a galaxy} may be sensitive 
to the selection of the inner boundary of the fit.

\citet{Meyer2013} determined the structural properties of the underlying host of BCDs and dEs in the Virgo Cluster
and also found the BCD hosts to be smaller than the dIrrs.
However, they argued in addition that the BCD hosts and dIrrs together 
are largely overlapping with the structural parameters of dEs.
Moreover, the overlapping will be more significant if the different distance of each BCD from the Virgo Cluster centre are taken into consideration.

Since {NIR} light traces the old stellar population, many authors have {used} deep {NIR imaging} to study the structural 
properties of the underlying {hosts} of dwarf galaxies
\citep{cairo2003,Micheva2013,janowiecki2014,janz2014,young2014}. 
However, the deepest {NIR images} (usually {$H$} band) only {reach} $\sim$ 23~mag ${\rm arcsec^{-2}}$  
for BCDs \citep{Micheva2013} and $\sim$ 24~mag ${\rm arcsec^{-2}}$ for other dwarf galaxies \citep{janz2014,young2014}. 
To probe the underlying host of BCDs at larger {radii and better} constrain their structural properties,
{deeper NIR images are} needed. The Cosmic Assembly Near-IR Deep Extragalactic Legacy Survey (CANDELS, \citealt{grogin2011,koekemoer2011})
obtained extremely deep {images} of galaxies with {the Wide Field Camera 3 (WFC3/IR) and the Advanced Camera for Surveys on 
the {\sl Hubble Space Telescope (HST)}}, 
which have {a full width at half-maximum (FWHM) of $\sim$ 0.13 arcsec at NIR}. The survey targets five sky regions. Among them, 
{the Great Observatories Origins Deep Survey North and South fields (GOODS-N and GOODS-S)} are two 
sky regions with deeper {NIR} imaging than the other three. This deep {NIR} photometry survey allows detailed analysis of 
{the} structural properties of dwarf galaxies outside our nearby Universe. In this work, we select {a} BCD sample from {the} 
CANDELS GOODS-N and GOODS-S {fields}
and study the structural properties of the
underlying {hosts} of these BCDs. 

{Throughout this paper, we adopt the cosmological parameters $H_0=70\, {\rm km s^{-1} Mpc}^{-1}$, $\Omega_{\Lambda}=0.73$ 
and $\Omega_{\rm m}=0.27$. 
All magnitudes in this paper are given in the AB photometric system unless otherwise stated.}

\section{Sample selection}

There are many definitions of BCDs {according to} their distinctive spectral features or morphological properties. The widely used definition in 
\citet{gilde2003} defines galaxies with {the} following properties as {BCDs: the galaxy is}
1) blue with $B-R < 1$; 2) compact with peak surface brightness {in the $B$} band 
$\mu_{\rm peak} < 22$~mag~${\rm arcsec}^{-2}$; 3) a dwarf with $M_{\rm K} > -21$~mag. 
In this work, we select BCDs {in the GOODS-N and GOODS-S fields} from the WFC3-selected photometry {catalogue} of \citet{skelton2014} and \citet{brammer2012}. 
We use the definition in \citet{gilde2003} 
but parametrize `dwarf' via the stellar mass {determined} by fitting the {spectral energy distributions}
from optical to {NIR}  \citep{skelton2014}. To obtain a BCD sample with robust stellar mass and physical size estimation, 
spectroscopic redshift is needed. Considering the redshift dimming effect, the low {surfaces} of galaxies are
not detectable at high redshift. Therefore, only galaxies {with} $z < 0.25$ are selected. Our final selection criteria {for} BCD are: 
1) peak surface brightness at {\sl HST} {$F435W$}
$\mu_{\rm peak} < 22$~mag~${\rm arcsec}^{-2}$; 2) restframe colour of $B-R < 1$; 3) stellar mass $M_* < 10^9\ M_{\odot}$; 4) redshift $z < 0.25$. 
Finally, 20 galaxies in the GOODS-N {field} and 14 in the GOODS-S field are {selected} as BCD {via} spectroscopic redshift measurement. 
{Stellar mass
and spectroscopic redshift are taken from the catalogue of \citet{skelton2014}, who estimate stellar mass by using {\textsc FAST} code \citep{kriek2009} and compile
spectroscopic redshift from literature.}
The {\sl HST} {$F160W$ and $F435W$} band {images} of these BCDs are shown in {Figs. 1 and 2} with {an} image size of 100 arcsec $\times$100 arcsec \citep{skelton2014}
\footnote{We use $F160W$ and $F435W$ mosaics from 3D-HST website
http://3dhst.research.yale.edu/Home.html}.

The basic properties of the BCD sample are listed in Table 1. We also calculate the rotational asymmetry of BCDs 
 by using the publicly available code, MORPHEUS \citep{abraham2007}. The asymmetry is measured
by rotating the galaxy image by $180\,^{\circ}$ and subtracting it from the original image \citep{conselice2003}.
For simplicity, {object IDs} in the GOODS-N field
are prefixed with `GN' and in the GOODS-S field with `GS'.

\begin{figure*}
 \centering
 \includegraphics[width=18cm]{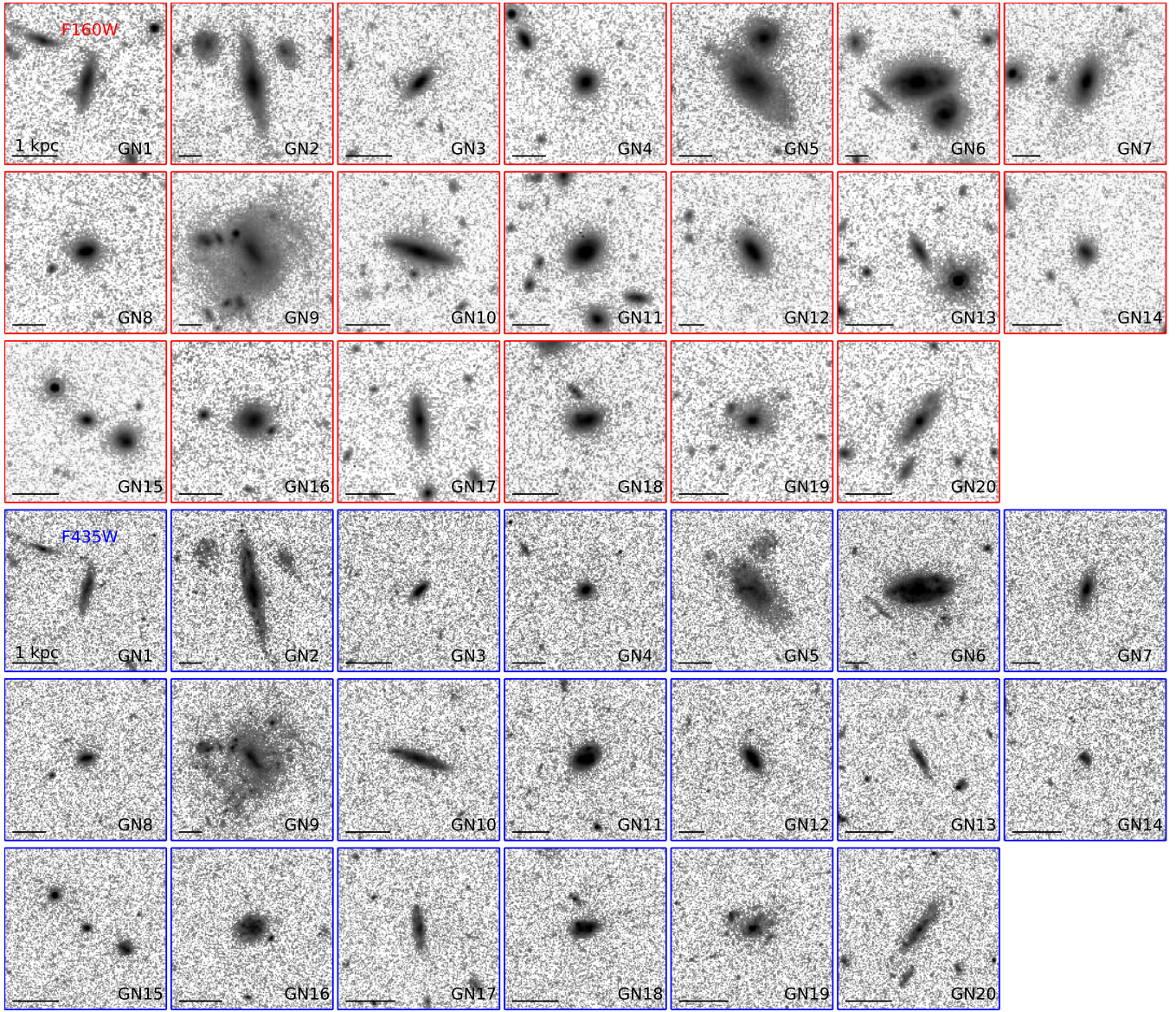}
 \caption{{$F435W$ and $F160W$} images of BCDs in the GOODS-N field {with logarithmic gray-scale}.
 Upper three rows are for BCDs in the $F160W$ band and bottom three rows for the $F435W$ band.} The image size is 100 arcsec $\times$100 arcsec. 
 {Length of the 
 solid line in the bottom left of each panel indicates
 the size of 1 kpc at the redshift of the BCD. 
 {$F160W$} images have red frames while {$F435W$} image frames are blue.
 Each panel is marked with the {object ID}.}
\label{figure1}
\end{figure*}

\begin{figure*}
 \centering
 \includegraphics[width=\textwidth]{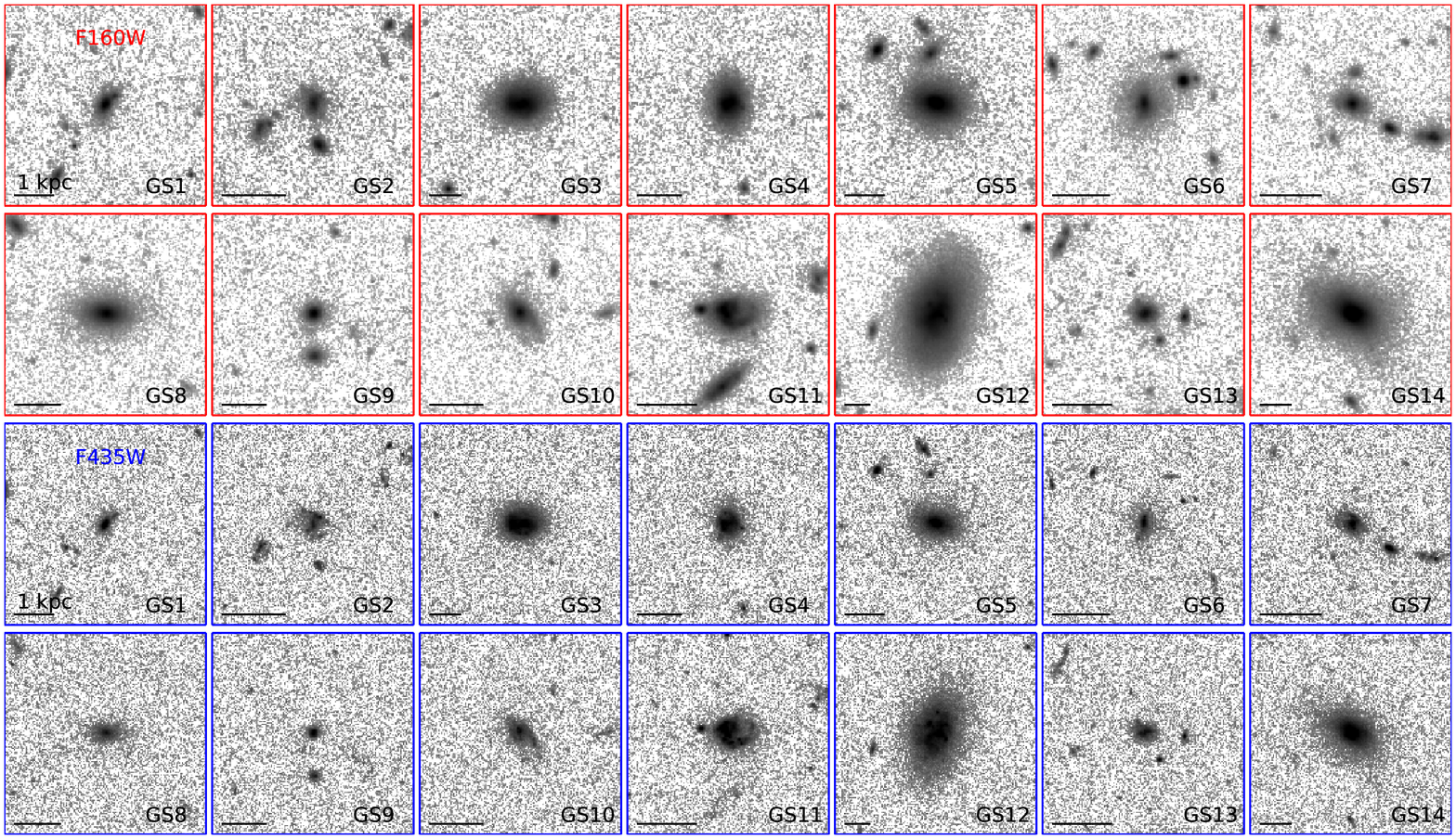}
 \caption{{Same} as Fig. 1 but with BCD sample in the GOODS-S field.}
\label{figure2}
\end{figure*}

\begin{table*}
  \caption{Basic properties of BCD sample.}
  \tabcolsep=0.16cm
  \label{table:1}
  \centering
  \begin{tabular}{l c c c c c c c c c}
\hline\hline
 ID & RA (J2000) & Dec. (J2000) & ${\rm m_{F435W}}$ & ${\rm m_{F160W}}$ & $z$ $^a$& ${\rm log(M_{star})}$ &
  $B-H$ $^b$ & Asym435 $^c$ & Asym160\\
  & (h:m:s) & ($^{\circ}$\ $'$\ $''$) & (mag) & (mag) & &(${\rm M_{\odot}})$ & (mag)  & \\
\hline
GN1  &  12:36:26.03 & 62:07:50.9 & 23.00$\pm$0.01 & 21.36$\pm$0.01 & 0.20 & 8.66 & 1.15 & 0.07 & 0.05\\
GN2  &  12:36:43.35 & 62:08:19.3 & 21.52$\pm$0.01 & 20.53$\pm$0.01 & 0.09 & 8.21 & 0.85 & 0.11 & 0.04\\
GN3  &  12:36:51.12 & 62:09:38.6 & 22.27$\pm$0.00 & 21.27$\pm$0.01 & 0.21 & 8.41 & 0.74 & 0.14 & 0.09\\
GN4  &  12:36:51.65 & 62:09:54.6 & 22.14$\pm$0.01 & 21.01$\pm$0.00 & 0.14 & 8.41 & 1.06 & 0.13 & 0.10\\
GN5  &  12:36:32.46 & 62:10:37.4 & 21.60$\pm$0.01 & 19.95$\pm$0.01 & 0.14 & 8.90 & 1.24 & 0.11 & 0.07\\ 
GN6  &  12:36:41.63 & 62:11:31.8 & 20.79$\pm$0.00 & 19.56$\pm$0.00 & 0.09 & 8.70 & 0.97 & 0.35 & 0.12\\
GN7  &  12:37:18.29 & 62:11:50.7 & 22.25$\pm$0.00 & 20.69$\pm$0.00 & 0.11 & 8.40 & 1.45 & 0.07 & 0.03\\
GN8  &  12:36:06.38 & 62:12:53.1 & 22.95$\pm$0.01 & 20.77$\pm$0.00 & 0.14 & 8.75 & 1.82 & 0.06 & 0.04\\
GN9  &  12:36:59.39 & 62:14:04.8 & 21.01$\pm$0.01 & 19.94$\pm$0.01 & 0.09 & 8.46 & 0.99 & 0.14 & 0.10\\
GN10 &  12:36:43.75 & 62:13:56.8 & 22.32$\pm$0.01 & 20.48$\pm$0.00 & 0.20 & 8.96 & 1.25 & 0.07 & 0.06\\
GN11 &  12:36:17.42 & 62:14:16.4 & 21.46$\pm$0.00 & 19.97$\pm$0.00 & 0.16 & 8.98 & 1.22 & 0.17 & 0.05\\
GN12 &  12:36:56.50 & 62:14:19.9 & 21.66$\pm$0.00 & 20.69$\pm$0.00 & 0.10 & 8.21 & 0.99 & 0.13 & 0.05\\
GN13 &  12:37:29.99 & 62:15:59.6 & 23.63$\pm$0.01 & 22.30$\pm$0.01 & 0.22 & 8.35 & 0.89 & 0.10 & 0.04\\
GN14 &  12:36:30.39 & 62:15:58.7 & 23.63$\pm$0.01 & 22.26$\pm$0.01 & 0.23 & 8.12 & 0.97 & 0.12 & 0.05\\
GN15 &  12:36:51.47 & 62:17:33.2 & 23.47$\pm$0.01 & 22.36$\pm$0.01 & 0.21 & 8.27 & 0.91 & 0.22 & 0.05\\
GN16 &  12:37:57.07 & 62:18:05.0 & 22.26$\pm$0.01 & 20.87$\pm$0.01 & 0.19 & 8.80 & 0.98 & 0.17 & 0.07\\
GN17 &  12:37:08.82 & 62:18:18.0 & 23.14$\pm$0.01 & 21.16$\pm$0.01 & 0.23 & 8.84 & 1.42 & 0.11 & 0.08\\
GN18 &  12:37:25.44 & 62:19:08.0 & 21.97$\pm$0.00 & 20.83$\pm$0.01 & 0.21 & 8.59 & 0.70 & 0.32 & 0.09 \\
GN19 &  12:37:21.25 & 62:19:15.5 & 22.61$\pm$0.01 & 21.24$\pm$0.01 & 0.23 & 8.47 & 0.84 & 0.15 & 0.08\\
GN20 &  12:37:21.71 & 62:20:05.3 & 23.31$\pm$0.01 & 21.22$\pm$0.01 & 0.21 & 8.80 & 1.29 & 0.11 & 0.07\\
\hline
GS1  &  03:32:36.98  & -27:55:24.1 & 22.56$\pm$0.00 & 21.74$\pm$0.01 & 0.13 & 8.04 & 0.75 & 0.18 & 0.10 \\
GS2  &  03:32:49.75  & -27:54:03.9 & 23.27$\pm$0.01 & 21.83$\pm$0.02 & 0.24 & 8.61 & 0.78 & 0.20 & 0.06\\
GS3  &  03:32:42.55  & -27:52:52.3 & 21.12$\pm$0.00 & 20.06$\pm$0.00 & 0.10 & 8.48 & 0.91 & 0.10 & 0.06\\
GS4  &  03:32:42.81  & -27:52:40.6 & 21.82$\pm$0.00 & 20.30$\pm$0.01 & 0.15 & 8.99 & 1.22 & 0.25 & 0.05\\
GS5  &  03:32:52.12  & -27.51:08.7 & 21.08$\pm$0.02 & 19.80$\pm$0.00 & 0.13 & 8.88 & 1.20 & 0.09 & 0.04\\
GS6  &  03:32:40.79  & -27:50:35.1 & 22.55$\pm$0.01 & 21.37$\pm$0.01 & 0.21 & 8.61 & 0.80 & 0.11 & 0.06\\
GS7  &  03:32:11.38  & -27:49:17.8 & 22.69$\pm$0.01 & 21.89$\pm$0.01 & 0.23 & 8.37 & 0.43 & 0.12 & 0.03\\
GS8  &  03:32:48.91  & -27:49:05.2 & 22.93$\pm$0.01 & 20.91$\pm$0.01 & 0.16 & 8.73 & 1.50 & 0.06 & 0.05\\
GS9  &  03:32:53.75  & -27:48:51.0 & 23.45$\pm$0.01 & 21.98$\pm$0.01 & 0.15 & 8.16 & 1.18 & 0.11 & 0.11\\ 
GS10 &  03:32:25.13  & -27:47:24.3 & 22.99$\pm$0.01 & 21.67$\pm$0.01 & 0.19 & 8.39 & 0.70 & 0.09 & 0.07\\
GS11 &  03:32:19.18  & -27:44:46.3 & 21.16$\pm$0.00 & 20.40$\pm$0.00 & 0.22 & 8.69 & 0.46 & 0.37 & 0.32\\
GS12 &  03:32:28.34  & -27:44:26.2 & 20.37$\pm$0.00 & 19.27$\pm$0.00 & 0.08 & 8.62 & 0.99 & 0.28 & 0.04\\
GS13 &  03:32:19.25  & -27:44:38.9 & 23.20$\pm$0.01 & 21.92$\pm$0.01 & 0.22 & 8.44 & 0.70 & 0.09 & 0.08\\
GS14 &  03:32:33.39  & -27:43:49.1 & 20.84$\pm$0.00 & 19.53$\pm$0.00 & 0.10 & 8.63 & 1.18 & 0.11 & 0.06\\
\hline
\end{tabular}\\
Notes:
$^a$ Redshift.\\
$^b$ Global $B-H$ colour using total magnitude.\\
$^c$ Asymmetry is given for both bands in the two last columns.\\

\end{table*}

\section{Surface Brightness Profiles and Fitting}

In this paper, we focus on the structural properties of underlying {hosts} of BCDs at {\sl HST} {$F435W$ and $F160W$} bands. 
We run the {Image Reduction and Analysis Facility ({\textsc IRAF}) Ellipse task} 
\citep{Jedrzejewski} 
to fit elliptical isophotes to the images and
measure the light profiles. {Ellipse}
fit the isophotes from the {centre} towards the outskirts {along the semimajor axes} with step size of 1 pixel.
{We use the {\textsc SEXTRACTOR} program \citep{bertin1996} 
to distinguish objects that do not belong to the BCD  
and then mask them out.
The geometric parameters (centre, ellipticity and position angle) are obtained by following the procedure in \citet{li2011}.
First, we run Ellipse and set all three geometry parameters free. The centre of the BCD 
is the average central position of ellipses within $\sim$ 0.3--0.6 arcsec. Then, we fix the centre to the value just
obtained and run Ellipse again, while the ellipticity and position angle are still set free. Typically we
calculate these two parameters as the average value at the outer regions of the galaxy, where the intensity 
is about three times above the sky. Finally, we run Ellipse to obtain the 
surface brightness profiles (SBPs), fixing the geometric parameters
to the values obtained above.}
We estimate the sky background noise by binning the sky background pixel (galaxies and stars are masked out) with binsize comparable to the 
size of the object and calculating the variance of values in all bins. The surface brightness uncertainties are {the} quadratic sum of those
in {the Ellipse} result {with} the sky background noise {just determined}.
A signal-to-noise ratio criterion (S\/N $>$ 3) 
{for} isophote is set to obtain the SBP.

The obtained SBPs are shown in {Figs. 3 and 4} for BCDs in the GOODS-N and GOODS-S {fields}, respectively.
{Ellipticity and position angle values are shown in the upper right corner.}
The squares in each panel
represent the SBPs {in the $F435W$} band and {the} circles represent SBPs {in the $F160W$} band. 
In the case of {the BCD} GS2, which has extremely irregular morphology {in the $F435W$} band,
no reliable SBP can be extracted from {the $F435W$} image.
The surface brightness is corrected for the redshift dimming effect by a factor of 
$(1+z)^4$. It can be seen that the surface brightness {in the $F160W$} band reaches $\sim$ 26~${\rm mag\ arcsec^2}$ which is $\sim$ 1.5 mag deeper than the 
deepest {NIR} photometry of BCDs. 
{The} CANDELS survey is much deeper than the previous {NIR} photometry of dwarf galaxies and can detect the outskirts at larger radius,
which may be essential to constrain the structural properties of the underlying old stellar population.
{We also plot $B-H$ colour versus radius in the bottom of each panel. 
The colour profiles are generally flat or ascending with radius.}

In some cases, there are some {bumps} in the SBPs.
{It can be seen that the SBP tends to be smoother in the $F160W$ than in the $F435W$ band. 
Meanwhile, asymmetry of BCDs at $F160W$ band are generally lower than that at $F435W$ band.
Therefore, these {bumps} in the SBPs are possibly due to distribution of star formation regions in the outskirts.
The asymmetry of BCDs in the $F435W$ band seems to be much lower than the massive starburst galaxies, such as
ultraluminous infrared galaxies \citep{conselice2003}.}
There are two emission lines from star forming regions 
that could have possibly affected the observed {$H$}-band photometry of our sample: \feii at 1.25~$\mu$m
and Pa~$\beta$ at 1.28~$\mu$m. However, the equivalent widths (EWs) of these two lines are generally low. For  
typical NIR spectra of H{\sc ii} galaxies 
\citep{martins2013} and irregular galaxies \citep{mannucci2001}, the EWs are less than 10 \AA. Besides, we {might} expect that the emission lines could be 
more weak in the outskirts of BCDs. Therefore, the SBP {in the $H$} band will not be significantly affected by star forming regions. 

\begin{figure*}
 \begin{center}
  \includegraphics[width=\textwidth]{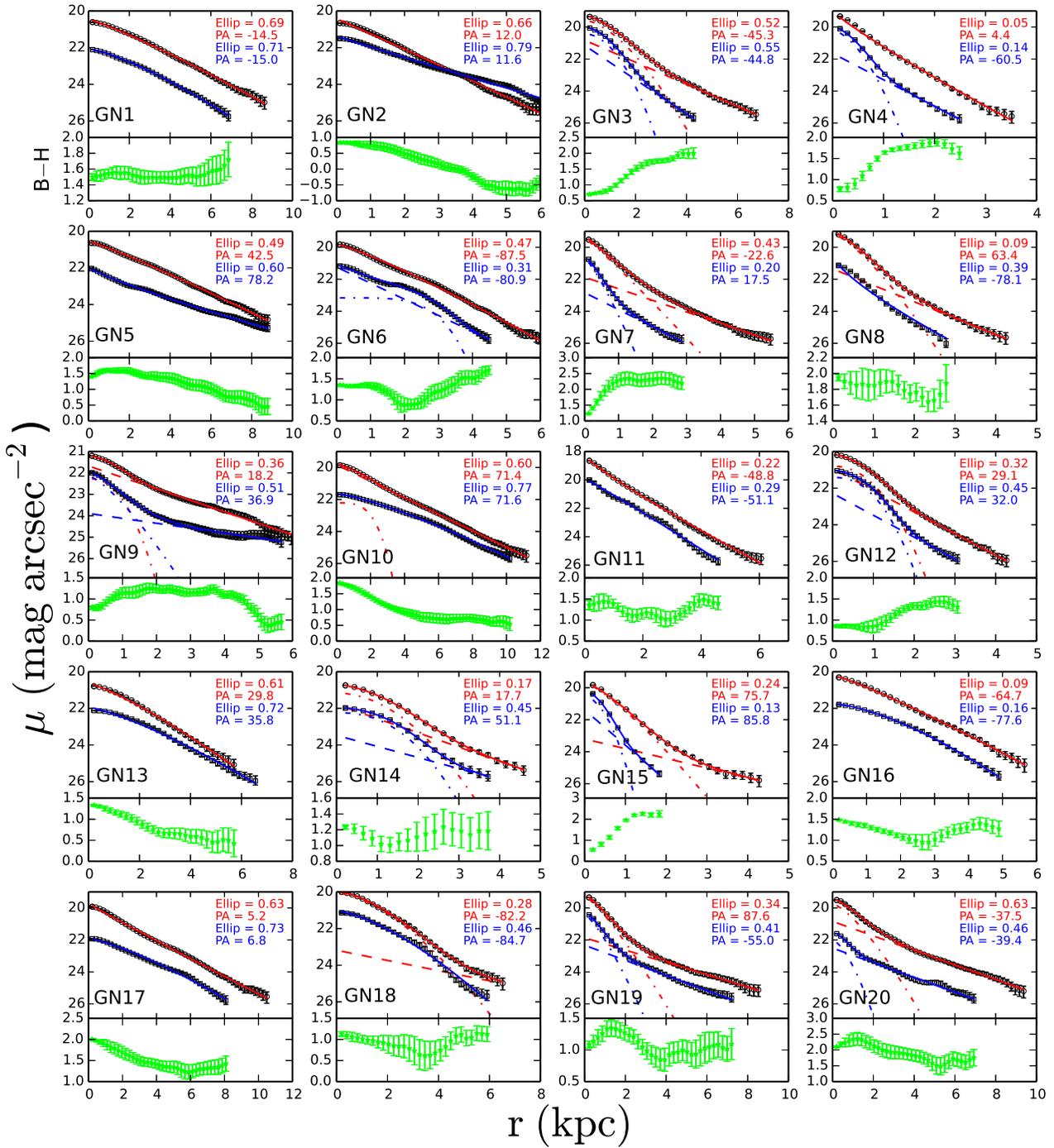}
 \end{center}
 \caption{{SBPs} of BCDs {in the $F160W$ (circles)} and {$F435W$ (squares)} bands. The best-fitting model is shown as solid lines. 
If the two-component model {are} used,  
the two components are shown as dashed and dot-dashed lines. 
{Some BCDs are well-fitted with one-component model and only solid lines are plotted in these panels.}
{The bottom of each panel shows 
$B-H$ colour profiles.} Each panel is marked with the {object ID}. 
Ellipticity and position angle values are shown in the upper right corner.
The upper two quantities are obtained in the $F160W$ band and the bottom two quantities with blue colour are for $F435W$ band.}
\label{figure3}
\end{figure*}

\begin{figure*}
 \begin{center}
  \includegraphics[width=\textwidth]{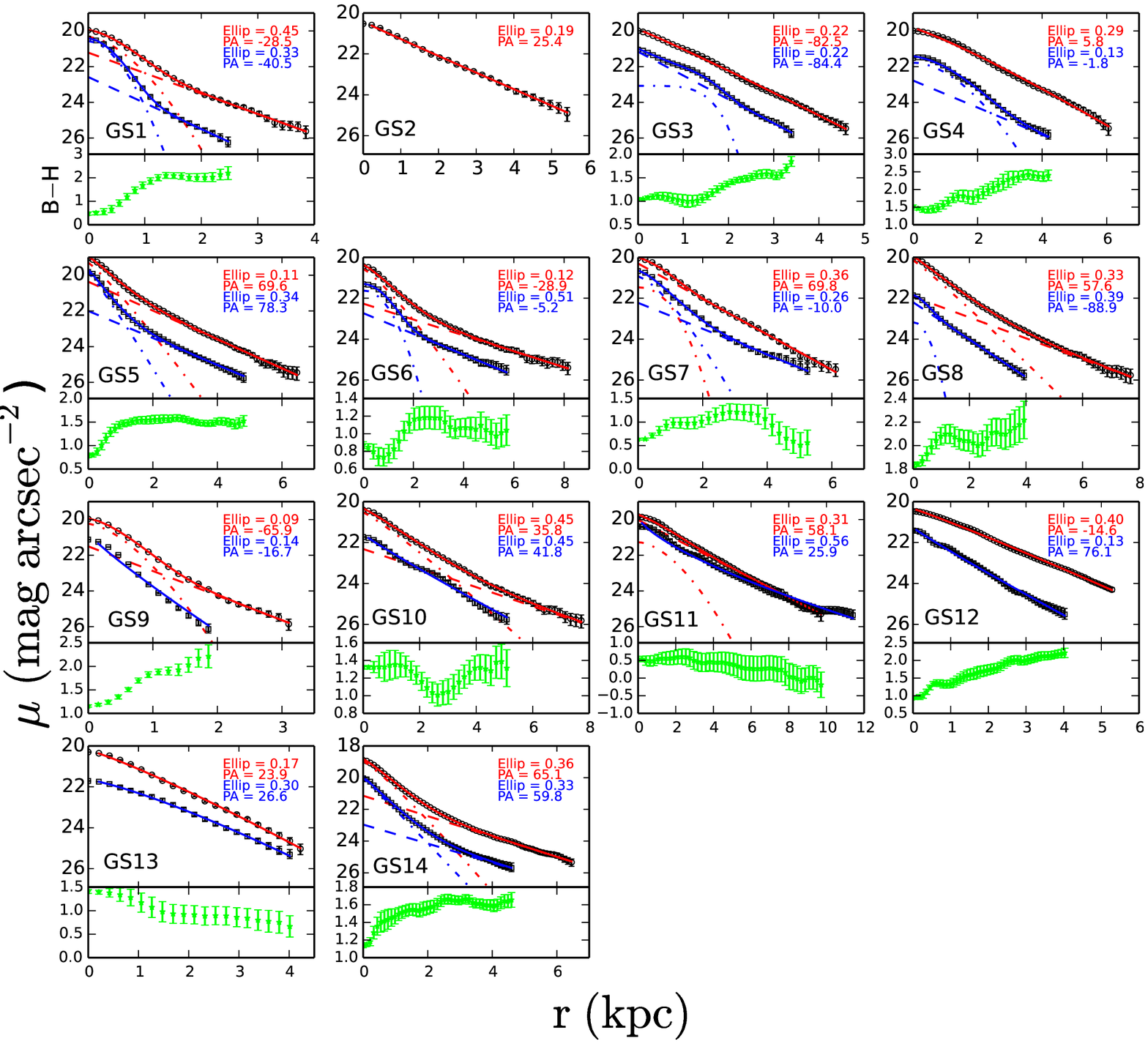}
 \end{center}
 \caption{{Same} as Fig. 3 but with BCD sample in the GOODS-S field.}
\label{figure4}
\end{figure*}

There are many functions that can be used to fit the {SBP of a} galaxy. The most widely-used is the S\'ersic function:
\begin{equation}
  I(r) = I_{\rm e}\ {\rm exp}(-b_n\ [(r/r_{\rm e})^{1/n}-1])
\end{equation}
with intensity $I_{\rm e}$ at $r_{\rm e}$. {The effective} radius $r_{\rm e}$ is the radius that encloses half of the total luminosity of the galaxy. 
The S\'ersic
index $n$ indicates the steepness of the profile. {Approximation} of the constant, $b_n$ as $1.9992n-0.3271$, is valid 
for $0.5<n<10$ \citep{prugenil1997}.
The {SBPs} of dwarf galaxies are usually fit with an exponential function (i.e., {$n=1$} in Eq. 1):
\begin{equation}
 I(r) = I_0\ {\rm exp}(-\frac{r}{h})
\end{equation}
with intensity $I_0$ at {the centre}, {and} scalelength $h$ 
(see \citealt{graham2005} for {the} transformation between scalelength $h$ and effective radius $r_{\rm e}$).

Two models are used to fit the SBPs: a one-component S\'ersic model and a two-component model combining an inner S\'ersic component
and an {outer} exponential component. 
{Although the two-component model is more flexible and always fits the SBPs better, 
we need to quantify the fit quality to quantitatively justify 
whether the two-component decomposition is indeed needed. 
The fit quality $Q$ is defined as the weighted quadratic sum of the difference between observed and best-fitting model SBP
\begin{equation}
 Q = \sum_{i} \frac{1}{\sigma_i \sum_{j} \frac{1}{\sigma_j}} (\mu_{\rm obs,i}-\mu_{\rm fit,i})^2 ,
\end{equation}
where   
$\sigma$ are the uncertainties of observed surface brightness, $\mu_{\rm obs}$. The summation is carried out over the SBP.
We fit the SBPs with the two models using the {\textsc IDL} routine MPFIT \citep{markwardt2009}. There are 
three free parameters in the S\'ersic function: surface brightness at effective radius} $\mu_{\rm e}$, S\'ersic index $n$ and effective radius $r_{\rm e}$. 
The initial values are arbitrarily chosen to be $\mu_{\rm e} = 21$~mag ${\rm arcsec}^{-2}$, $n = 1$ and $r_{\rm e} = 10$ pixels.
The surface brightness uncertainties are used for weighting with weight values of $1/\sigma^2$.

For {the $F435W$} band, we fit the {SBPs} of BCDs with {both the one- and two-component models. For 
the two-component model, the} blue dot-dashed and dashed lines in {Figs. 3 and 4}
represent the S\'ersic and exponential
components, respectively. The blue solid lines {represent their sum}.
The best-fitting parameters {are listed in Table 2. The `$Q$' column shows fit quality result,
with a lower value of this quantity suggesting a better fit. 
It is obvious that the two-component model yields a `better fit quality' in most cases. However, in some cases, where the fit quality of
the one-component model is relatively high or the number of data points is limited, the two-component model will yield 
unphysical parameter values in either the S\'ersic or the exponential component.
Moreover, the improvement in fit quality
using the two-component model is always limited in these cases. 
For 10 BCDs in GOODS-N and 7 BCDs in GOODS-S, we use the parameters of the one-component model for comparison.
The preference of decomposition is listed in the last column of Table 2.} 
The absolute magnitude is calculated by integrating the best-fitting profile from {centre to infinity} \citep{young2014}. 
The {$F435W$ and $F160W$} bands trace
the young and old stellar population of galaxies, respectively, and roughly correspond to the
$B$ and $H$ bands of the ground-based photometry system. We do $k$-corrections {for $F435W$ and $F160W$} bands by applying a correction factor of the 
difference between the observed {$F435W$ or $F160W$ magnitude and the} restframe $B$ or $H$ band magnitude. {These magnitudes are all
taken from} the catalogue of \citet{skelton2014}. 
{It should be noted that the spatial resolution is higher in the $F435W$ band images
(FWHM of $\sim$ 0.06 arcsec) than in the $F160W$ band images (FWHM of $\sim$ 0.15 arcsec).}
{To explore the resolution effect, we impose an additional Gaussian filter to the $F435W$ band images and reduce the resolution 
to the level of the $F160W$ band images. Then we obtain the structural parameters by the same procedure described above. 
The discrepancy between structural parameters of low and high resolution images is generally small.}

{For the $F160W$ band, both the one- and two-component models are used for fitting and comparison with other dwarf galaxies.
The best-fitting parameters are listed in Table 3.
The dot-dashed and dashed red lines in Figs. 3 and 4 show the S\'ersic and exponential components
in the two-component fitting. The red solid lines represent their sum. It can be seen that for the most part 
the two-component model fits the SBP of BCDs better than the one-component model. 
For 10 BCDs in GOODS-N and 5 in GOODS-S we use the parameters of the one-component model, which is more robust than 
the two-component model, for comparison at $H$ band.} 
{\citet{janowiecki2014} fit the outer region of BCDs with an exponential function. 
They argued that the outer regions of BCDs have colours consistent with evolved stellar populations by comparing the $B-H$ colour of the outer region
with simple stellar population models. For BCDs with moderately low metallicities, the outer regions have colours 
at $\sim$ 2.5 mag in the Vega photometric system. It can be seen in the colour profiles in Figs. 3 and 4 that, for the outer regions of BCDs where 
the exponential component dominates the SBP, the $B-H$ colour is almost redder than 1.0 mag, which correspond to $\sim$ 2.3 mag
in the Vega photometric system. Therefore, the outer regions of BCDs are generally dominated by evolved stellar populations 
and the exponential component mainly reflects the SBP of the underlying host of the BCD.
However, the inner S\'ersic component may not correspond to the star forming region at the centre.}

\section{Comparison with other dwarf galaxies}

In Fig. 5, we compare the structural parameters of our BCDs to those for the dwarf galaxies in the literature in the $B$ band. 
To be consistent
with previous works where only the parameters for the outer regions of galaxies are measured, we use the 
exponential components of our BCDs for comparison.
As discussed in Section3, the parameters of the one-component model are used instead for 17 BCDs 
where that model is sufficient.
Blue circles represent BCDs at a redshift of 0.09--0.15 and are marked as `BCDz1'. 
Red circles show BCDs at redshift of 0.15--0.25 and are marked as `BCDz2'.
Although BCDs at the higher redshift range tend to be brighter
which could possibly be due to {a} selection effect, all BCDs at the two redshift ranges are {on} the same sequence of luminosity-radius 
{relationship}.
We include the structural parameters of dwarf galaxies {in the $B$} band compiled from {the} literature in the comparison.
Black crosses in Fig. 5 are BCDs from \citet{marlowe1997} and \citet{cairo2001}. {Purple} triangles are dIrrs
from \citet{patterson1996} and \citet{vanzee2000} and {green} squares are dEs from \citet{binggeli1993}.
{The magnitudes in the literature are converted from the Vega to the AB photometric system with a conversion factor of $-$0.163 mag 
in the $B$ band.}
{The} absolute magnitude in Fig. 5 represents the total luminosity of {the} BCDs. 
{It can be seen that} 
the underlying {hosts} of our BCDs are consistent with the BCDs in \citet{cairo2001} and \citet{marlowe1997} in the 
luminosity-radius diagram.  
While the {effective radius} of the underlying host of our BCDs
is immensely small compared to {dIrrs and dEs},
the central brightness is fainter than other BCDs but consistent with that of dIrrs and dEs.   

\begin{figure*}
 \centering
 \includegraphics[width=\textwidth]{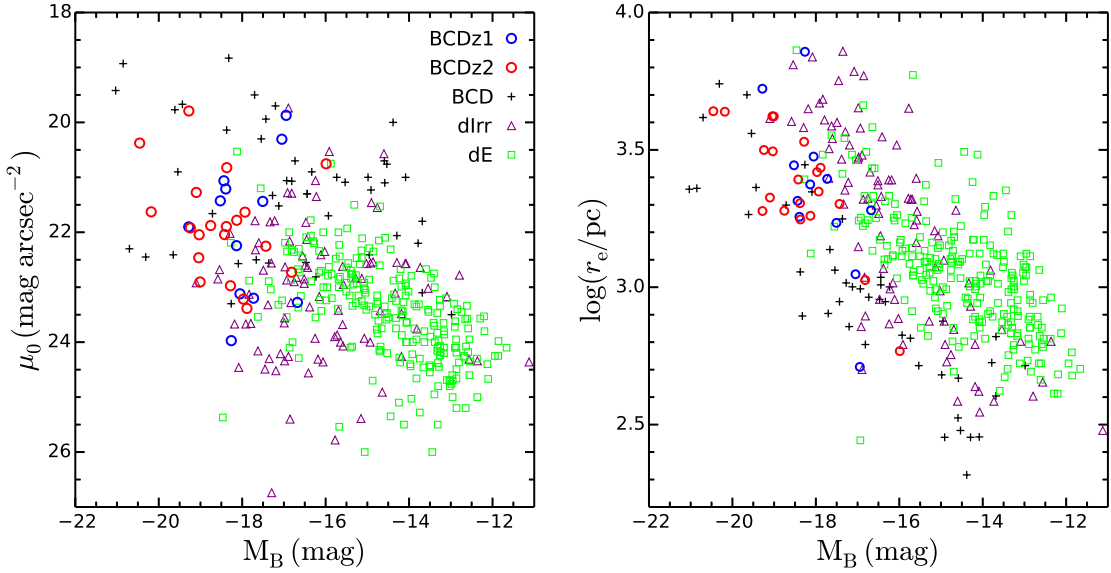}
 \caption{Comparison of structural parameters {in the $B$} band between the underlying {hosts} of BCDs and 
other dwarf galaxies. The blue circles represent our BCDs at redshift of 0.09 -- 0.15 and are marked
as `BCDz1'.
The red circles are our BCDs at redshift of 0.15 -- 0.25 and marked as `BCDz2'. 
The black crosses are the BCDs from \citet{cairo2001} and \citet{marlowe1997}. 
Data for dIrrs are compiled from \citet{patterson1996}, \citet{vanzee2000} and dEs from \citet{binggeli1993}.
}
\label{figure5}
\end{figure*}

{A comparison} of the {structural} 
parameters of BCDs with other dwarf galaxies {at $H$ bands} is shown in Fig. 6. 
The upper row is for the 
inner S\'ersic component and {the} 
bottom row for the outer exponential component.
 The empty red circles are our BCDs with one-component decomposition in this work and marked with `BCD1'.
The solid red circles are 
BCDs with two-component decomposition and marked with `BCD2'.
{\citet{janz2014} fit two-dimensional models to the images of dEs in Virgo cluster with both a one-component S\'ersic model
and a two-component model combining an inner S\'ersic and an outer exponential component.}\
{About one third of the dEs in \citet{janz2014} favour the simple one-component decomposition. 
The parameters of the one-component model for these dEs are also included in the comparison of underlying hosts in the bottom row of Fig. 6.}
The empty green squares represent dEs in {the} Virgo Cluster from \citet{janz2014} with one-component decomposition and 
solid green squares for dEs with two-component decomposition.
{Although \citet{noeske2003,noeske2005} only fit the outer regions of BCDs with an exponential function,
their results should be comparable to the outer exponential component of the two-component decomposition.}
Therefore, we include these BCDs in the comparison of the {exponential component in the bottom row of} Fig. 6.
{Since dIrrs always have approximately exponential SBPs, we also include parameters for dIrrs for the comparison of underlying 
hosts in the bottom row in Fig. 6.} 
{Purple triangles represent} dIrrs
from \citet{young2014} and \citet{mccall2012}, {who fit the whole galaxy with an exponential function.}
Since the parameters in \citet{mccall2012} were derived {in the} $K_s$ band,
we use a colour of $H-K=0.1$ to obtain the equivalent surface brightness and magnitude at $H$ band.
{We convert the supplied Vega magnitudes in literature to the AB photometric system with a conversion factor of 1.37 mag in the $H$ band.}

{It should be noted that 
the absolute magnitudes of the inner and outer components are calculated by integrating the respective best-fitting profile of each component.
This is different from the comparison in Fig. 5 where we used the absolute magnitude of the whole galaxy. 
It is interesting to note that the outer components of BCDs are in remarkable agreement with those of dEs. 
Their underlying hosts show a similar S\'ersic index, extrapolated central surface brightness and effective radius.
Moreover, all the outer regions of dwarf galaxies seem to follow a consistent luminosity-size relation, 
though the dIrrs tend to be slightly larger than BCDs at the 
low-luminosity end.}
\citet{janowiecki2014} found that the underlying hosts of BCDs are significantly smaller and {centrally} brighter than dIrrs.
With more dIrrs in {our} comparison, 
the difference in structural parameters between BCDs and dIrrs 
seems to be less significant {in the $H$ band than in the $B$ band}.
Unlike the outer {components}, the inner components of BCDs seem to {have lower S\'ersic index $n$ and are} 
fainter in the
{centre} compared to dEs. {Nevertheless, the} size of the inner {components of BCDs}
is consistent with {those of dEs} at a fixed luminosity.

\begin{figure*}
 \centering
 \includegraphics[width=15cm]{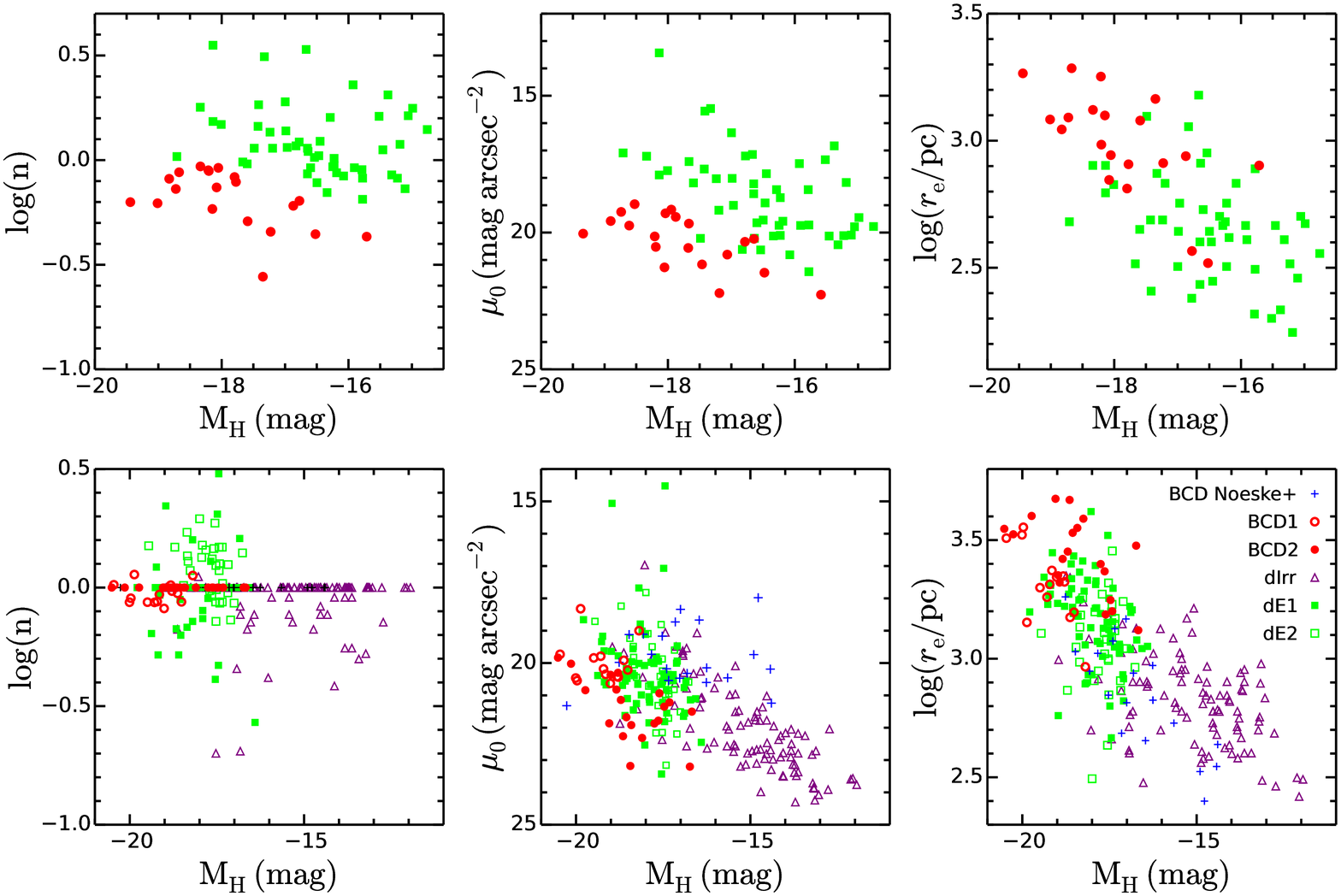}
\caption{Comparison of structural parameters {in the $H$} band
of BCDs and other type dwarf galaxies. 
The empty red circles are BCDs with one-component decomposition in this work and marked with `BCD1'.
The solid red circles are 
BCDs with two-component decomposition and marked with `BCD2'.
The blue crosses are BCDs from \citet{noeske2003,noeske2005} and {purple}
triangles are dIrrs from \citet{young2014} and \citet{mccall2012}. 
The empty green squares represent dEs in {the} Virgo Cluster from \citet{janz2014} with one-component decomposition and marked
with `dE1'.
The filled green squares are dEs with two-component decomposition and marked with `dE2'.
The parameters of inner {S\'ersic} component 
are shown in the upper row
and outer {exponential} component in bottom row.}
\label{figure6}
\end{figure*}

\begin{table*}
 \caption{Parameters from SBP fitting {in the $F435W$}.}
 \tabcolsep=0.11cm
 \centering
  \begin{tabular}{l|ccccc|c|cccc|c|ccccc|c}
  \toprule
  & \multicolumn{5}{c}{One} & \mc{} & \multicolumn{4}{|c}{Inner} & \mc{} &
    \multicolumn{5}{|c|}{Outer}  \\
  & \multicolumn{5}{c}{component} & \mc{} & \multicolumn{4}{|c|}{component} & \mc {} &
    \multicolumn{5}{|c|}{component} & \mc{${\rm Preference}^c$}\\
    \cmidrule{2-6} \cmidrule{7-11} \cmidrule{12-17} 
    \multicolumn{1}{c|}{ID} & \mc{$m$} & \mc{$n$} & \mc{$\mu_{435}$} & \mc{$r_{\rm e}^a$} & \multicolumn{1}{c|}{$Q^b$} & & 
              \mc{$m$} & \mc{$n$} & \mc{$\mu_{435}$} & \multicolumn{1}{c|}{$r_{\rm e}$} & & 
              \mc{$m$} & \mc{$n$} & \mc{$\mu_{435}$} & \mc{$r_{\rm e}$} & \multicolumn{1}{c|}{$Q$}  \\
    & \mc{(mag)} & &  \mc{(mag ${\rm arcsec^{-2}}$)} & \mc{(kpc)} & & &
    \mc{(mag)} & & \mc{(mag ${\rm arcsec^{-2}}$)} & \mc{(kpc)} & &
    \mc{(mag)} & & \mc{(mag ${\rm arcsec^{-2}}$)} & \mc{(kpc)} & & \\        
  \midrule
 GN1  & -19.0 & 0.78 & 22.7 & 3.1 & 0.08 & & -4.1  & 0.20 & 30.0 & 0.1 & & -19.0 & 1.00 & 22.4 & 3.3 & 0.11 & one\\
 GN2  & -18.5 & 1.05 & 21.7 & 2.8 & 0.07 & & -14.4 & 0.20 & 24.0 & 0.6 & & -18.5 & 1.00 & 21.8 & 2.8 & 0.05 & one\\
 GN3  & -18.7 & 1.09 & 20.4 & 1.3 & 0.23 & & -18.2 & 0.57 & 21.0 & 0.9 & & -17.8 & 1.00 & 22.0 & 1.9 & 0.07 & two\\
 GN4  & -16.9 & 1.34 & 19.9 & 0.7 & 0.26 & & -15.2 & 0.56 & 20.6 & 0.4 & & -16.7 & 1.00 & 22.0 & 1.0 & 0.05 & two\\
 GN5  & -19.3 & 1.26 & 22.4 & 5.3 & 0.06 & & -14.4 & 0.55 & 23.8 & 0.6 & & -19.2 & 1.00 & 22.8 & 5.0 & 0.04 & one\\
 GN6  & -18.4 & 0.62 & 21.7 & 1.8 & 0.16 & & -17.1 & 0.20 & 23.5 & 1.8 & & -17.9 & 1.00 & 21.6 & 1.8 & 0.07 & two\\
 GN7  & -16.6 & 1.71 & 20.5 & 1.1 & 0.29 & & -15.8 & 0.84 & 21.0 & 0.5 & & -16.0 & 1.00 & 23.6 & 1.9 & 0.08 & two\\
 GN8  & -17.0 & 1.50 & 20.8 & 1.1 & 0.27 & & -16.7 & 0.87 & 21.3 & 0.7 & & -16.3 & 1.00 & 25.0 & 3.5 & 0.08 & one\\
 GN9  & -18.3 & 2.15 & 21.7 & 6.5 & 0.22 & & -15.8 & 0.79 & 22.4 & 0.9 & & -18.1 & 1.00 & 24.2 & 7.2 & 0.05 & two\\
 GN10 & -20.2 & 0.78 & 22.4 & 4.4 & 0.05 & & -4.2  & 0.20 & 30.0 & 0.1 & & -20.2 & 1.00 & 22.0 & 4.4 & 0.18 & one\\
 GN11 & -19.3 & 1.24 & 20.3 & 1.9 & 0.14 & & -16.2 & 0.35 & 21.2 & 0.4 & & -19.2 & 1.00 & 20.9 & 2.1 & 0.11 & one\\
 GN12 & -17.6 & 1.11 & 20.9 & 1.3 & 0.29 & & -17.2 & 0.52 & 21.5 & 1.0 & & -16.6 & 1.00 & 23.3 & 2.5 & 0.04 & two\\
 GN13 & -18.4 & 0.89 & 22.7 & 2.5 & 0.16 & & -18.2 & 0.61 & 22.9 & 2.0 & & -18.9 & 1.00 & 26.6 & 20.5 & 0.05 & one\\
 GN14 & -17.8 & 0.86 & 22.5 & 1.7 & 0.13 & & -17.1 & 0.49 & 23.1 & 1.3 & & -17.1 & 1.00 & 23.9 & 2.7 & 0.04 & two\\
 GN15 & -16.7 & 1.21 & 20.2 & 0.5 & 0.25 & & -16.3 & 0.69 & 20.5 & 0.4 & & -15.7 & 1.00 & 23.0 & 1.1 & 0.08 & two\\
 GN16 & -18.4 & 0.67 & 22.5 & 2.0 & 0.03 & & -17.6 & 0.44 & 23.4 & 1.9 & & -17.7 & 1.00 & 23.0 & 2.4 & 0.03 & one\\
 GN17 & -19.2 & 0.81 & 22.7 & 3.2 & 0.09 & & -17.7 & 0.20 & 25.6 & 3.5 & & -18.9 & 1.00 & 22.6 & 2.9 & 0.09 & one\\
 GN18 & -19.1 & 0.72 & 21.8 & 2.1 & 0.16 & & -19.0 & 0.48 & 22.1 & 1.9 & & -19.2 & 1.00 & 26.2 & 20.4 & 0.12 & one\\
 GN19 & -19.2 & 2.29 & 20.1 & 3.6 & 0.28 & & -18.5 & 1.40 & 20.5 & 1.4 & & -18.9 & 1.00 & 25.0 & 9.8 & 0.13 & two\\
 GN20 & -19.2 & 1.77 & 21.8 & 4.3 & 0.16 & & -16.5 & 0.82 & 22.3 & 0.7 & & -18.9 & 1.00 & 23.4 & 4.2 & 0.08 & two\\
 \midrule
 GS1  & -16.2 & 1.47 & 20.3 & 0.7 & 0.37 & & -15.5 & 0.53 & 21.2 & 0.4 & & -15.2 & 1.00 & 23.2 & 1.2 & 0.08 & two \\
 GS2  & -     & -    & -    & -   & -    & & -     & -    & -    & -   & & -     & -    & -    & -   & -    & -\\
 GS3  & -18.4 & 0.96 & 21.3 & 2.1 & 0.07 & & -18.3 & 0.81 & 21.5 & 1.9 & & -17.2 & 1.00 & 26.1 & 11.0 & 0.06 & one \\
 GS4  & -17.9 & 0.88 & 21.8 & 1.6 & 0.15 & & -17.4 & 0.59 & 22.2 & 1.3 & & -16.9 & 1.00 & 23.7 & 2.6 & 0.07 & two\\
 GS5  & -18.1 & 1.83 & 19.6 & 1.6 & 0.16 & & -17.1 & 1.02 & 20.0 & 0.6 & & -17.6 & 1.00 & 22.5 & 2.4 & 0.07 & two\\
 GS6  & -18.3 & 1.61 & 21.5 & 2.6 & 0.27 & & -16.8 & 0.50 & 22.2 & 0.8 & & -18.0 & 1.00 & 23.5 & 3.4 & 0.06 & two\\
 GS7  & -18.4 & 1.22 & 21.2 & 1.8 & 0.16 & & -17.3 & 0.42 & 24.6 & 2.8 & & -17.8 & 1.00 & 21.3 & 1.2 & 0.14 & one\\
 GS8  & -17.5 & 1.34 & 22.1 & 1.9 & 0.09 & & -14.5 & 0.70 & 22.9 & 0.4 & & -17.4 & 1.00 & 22.9 & 2.0 & 0.05 & two\\ 
 GS9  & -16.0 & 1.09 & 21.2 & 0.6 & 0.16 & & -13.2 & 0.20 & 27.3 & 1.3 & & -15.9 & 1.00 & 21.2 & 0.5 & 0.13 & one\\
 GS10 & -17.9 & 1.22 & 22.3 & 2.2 & 0.14 & & -14.8 & 0.23 & 23.6 & 0.5 & & -17.8 & 1.00 & 22.8 & 2.3 & 0.08 & one\\
 GS11 & -20.5 & 1.39 & 20.8 & 4.4 & 0.17 & & -20.2 & 1.02 & 21.1 & 3.3 & & -20.0 & 1.00 & 25.4 & 21.2 & 0.10 & one\\
 GS12 & -17.5 & 1.01 & 21.6 & 1.7 & 0.10 & & -17.3 & 0.81 & 21.8 & 1.5 & & -17.0 & 1.00 & 25.7 & 9.0 & 0.09 & one\\
 GS13 & -18.1 & 0.82 & 22.4 & 1.8 & 0.05 & & -18.1 & 0.82 & 22.4 & 1.8 & & -10.0 & 1.00 & 30.0 & 1.6 & 0.05 & one\\
 GS14 & -18.0 & 1.69 & 19.7 & 1.4 & 0.20 & & -17.5 & 1.06 & 20.2 & 0.8 & & -17.1 & 1.00 & 23.4 & 3.0 & 0.05 & two\\
\bottomrule
  \end{tabular}\\
Notes:
$^a${All the surface brightness and radii correspond to the semi-major axes.}\\
$^b${fit quality: weighted quadratic sum of difference between observed and best-fitting model SBP.}\\
$^c${Preference for one-component model or two-component model.}\\
\end{table*}

\begin{table*}
 \caption{Parameters from SBP fitting {in the $F160W$}.}
 \tabcolsep=0.11cm
 \centering
  \begin{tabular}{l|ccccc|c|cccc|c|ccccc|c}
  \toprule
  & \multicolumn{5}{c}{One} & \mc{} & \multicolumn{4}{|c}{Inner} & \mc{} &
    \multicolumn{5}{|c|}{Outer} \\
  & \multicolumn{5}{c}{component} & \mc{} & \multicolumn{4}{|c|}{component} & \mc{} &
    \multicolumn{5}{|c|}{component} & \mc{Preference}\\
    \cmidrule{2-6} \cmidrule{7-11} \cmidrule{12-17}
    \multicolumn{1}{c|}{ID} & \mc{$m$} & \mc{$n$} & \mc{$\mu_{160}$} & \mc{$r_{\rm e}$} & \mc{$Q$} & &
              \mc{$m$} & \mc{$n$} & \mc{$\mu_{160}$} & \multicolumn{1}{c|}{$r_{\rm e}$} & &
              \mc{$m$} & \mc{$n$} & \mc{$\mu_{160}$} & \mc{$r_{\rm e}$} & \mc{$Q$} \\
    & \mc{(mag)} & & \mc{(mag ${\rm arcsec^{-2}}$)} & \mc{(kpc)} & & &
      \mc{(mag)} & & \mc{(mag ${\rm arcsec^{-2}}$)} & \mc{(kpc)} & &
       \mc{(mag)} & & \mc{(mag ${\rm arcsec^{-2}}$)} & \mc{(kpc)} &  \\      
  \midrule
 GN1  & -19.8 & 0.87 & 21.3 & 3.3 & 0.06 & & -19.3 & 0.68 & 21.7 & 2.7 & & -19.0 & 1.00 & 22.8 & 5.1  & 0.02 & one\\
 GN2  & -18.7 & 0.97 & 20.8 & 2.1 & 0.10 & & -18.0 & 0.73 & 21.4 & 1.6 & & -18.0 & 1.00 & 22.2 & 2.9  & 0.05 & one\\
 GN3  & -19.5 & 1.20 & 19.7 & 1.8 & 0.23 & & -18.9 & 0.62 & 20.4 & 1.2 & & -18.7 & 1.00 & 21.6 & 2.6  & 0.02 & two\\
 GN4  & -18.1 & 1.12 & 19.6 & 0.9 & 0.12 & & -18.0 & 0.96 & 19.7 & 0.9 & & -16.1 & 1.00 & 25.4 & 5.1  & 0.09 & one\\
 GN5  & -19.8 & 0.90 & 21.1 & 3.6 & 0.04 & & -19.5 & 0.82 & 21.4 & 3.2 & & -18.6 & 1.00 & 23.1 & 5.3  & 0.03 & one\\
 GN6  & -19.2 & 0.87 & 20.2 & 1.8 & 0.06 & & -1.5 & 0.24 & 30.0 & 0.0  & & -19.2 & 1.00 & 19.9 & 1.8  & 0.16 & one\\
 GN7  & -18.4 & 1.50 & 19.4 & 1.4 & 0.24 & & -17.9 & 0.92 & 19.9 & 0.9 & & -17.6 & 1.00 & 22.3 & 2.5  & 0.06 & two\\
 GN8  & -18.3 & 1.34 & 19.1 & 1.0 & 0.26 & & -17.9 & 0.74 & 19.7 & 0.7 & & -17.3 & 1.00 & 21.9 & 1.8  & 0.03 & two\\
 GN9  & -18.5 & 1.28 & 21.4 & 3.3 & 0.10 & & -15.6 & 0.43 & 22.6 & 0.8 & & -18.4 & 1.00 & 22.1 & 3.4  & 0.04 & two\\
 GN10 & -20.4 & 1.03 & 20.4 & 3.4 & 0.08 & & -17.2 & 0.28 & 23.0 & 1.5 & & -20.4 & 1.00 & 20.6 & 3.5  & 0.05 & two\\
 GN11 & -19.7 & 1.13 & 19.0 & 1.4 & 0.10 & & -19.7 & 0.97 & 19.2 & 1.4 & & -18.1 & 1.00 & 26.1 & 16.4 & 0.05 & one\\
 GN12 & -17.9 & 1.09 & 20.2 & 1.2 & 0.23 & & -17.1 & 0.45 & 21.2 & 0.8 & & -17.4 & 1.00 & 21.4 & 1.5  & 0.03 & two\\
 GN13 & -18.8 & 0.82 & 21.5 & 2.2 & 0.07 & & -18.4 & 0.63 & 21.8 & 1.8 & & -17.9 & 1.00 & 23.2 & 3.7  & 0.03 & one\\
 GN14 & -18.2 & 0.92 & 21.4 & 1.6 & 0.15 & & -17.5 & 0.51 & 22.1 & 1.2 & & -17.5 & 1.00 & 22.7 & 2.3  & 0.03 & two\\
 GN15 & -17.9 & 1.34 & 20.1 & 1.1 & 0.27 & & -17.7 & 0.79 & 20.5 & 0.8 & & -16.6 & 1.00 & 24.0 & 3.0  & 0.05 & two\\
 GN16 & -19.1 & 0.87 & 20.9 & 2.1 & 0.05 & & -18.9 & 0.85 & 21.0 & 2.0 & & -16.6 & 1.00 & 24.1 & 3.1  & 0.04 & one\\
 GN17 & -20.2 & 1.03 & 20.6 & 3.2 & 0.09 & & -16.7 & 0.24 & 23.3 & 1.2 & & -20.2 & 1.00 & 20.8 & 3.3  & 0.07 & one\\
 GN18 & -19.5 & 0.90 & 20.6 & 2.2 & 0.13 & & -19.3 & 0.63 & 20.9 & 1.8 & & -18.3 & 1.00 & 24.0 & 6.5  & 0.03 & two\\
 GN19 & -19.5 & 1.83 & 19.5 & 2.7 & 0.31 & & -18.7 & 0.81 & 20.1 & 1.1 & & -19.0 & 1.00 & 22.8 & 4.7  & 0.04 & two\\
 GN20 & -20.0 & 1.49 & 19.9 & 3.1 & 0.19 & & -18.6 & 0.73 & 20.6 & 1.2 & & -19.6 & 1.00 & 21.7 & 4.0  & 0.05 & two\\
 \midrule
 GS1  & -17.7 & 1.28 & 20.1 & 1.2 & 0.20 & & -16.8 & 0.60 & 20.9 & 0.9 & & -17.2 & 1.00 & 21.7 & 1.6 & 0.03 & two\\
 GS2  & -18.7 & 1.02 & 21.4 & 2.2 & 0.05 & & -15.0 & 0.23 & 24.1 & 0.1 & & -18.7 & 1.00 & 21.5 & 2.3 & 0.03 & one\\
 GS3  & -18.5 & 0.95 & 20.3 & 1.5 & 0.05 & & -16.0 & 0.20 & 23.4 & 0.2 & & -18.4 & 1.00 & 20.4 & 1.5 & 0.03 & one\\
 GS4  & -19.3 & 0.87 & 20.4 & 2.0 & 0.10 & & -19.3 & 0.87 & 20.4 & 2.0 & & -10.1 & 1.00 & 30.0 & 2.6 & 0.10 & one\\
 GS5  & -19.3 & 1.38 & 19.2 & 1.7 & 0.13 & & -18.0 & 0.89 & 19.8 & 1.0 & & -18.8 & 1.00 & 20.9 & 2.2 & 0.05 & two\\
 GS6  & -18.9 & 1.63 & 20.8 & 3.3 & 0.21 & & -17.7 & 0.83 & 21.4 & 0.6 & & -18.5 & 1.00 & 23.1 & 4.7 & 0.05 & two\\
 GS7  & -18.9 & 1.11 & 20.7 & 1.9 & 0.11 & & -16.5 & 0.44 & 22.4 & 0.3 & & -18.8 & 1.00 & 21.2 & 2.1 & 0.04 & two\\
 GS8  & -19.0 & 1.37 & 20.3 & 2.3 & 0.16 & & -18.2 & 0.93 & 20.8 & 1.3 & & -18.3 & 1.00 & 22.6 & 3.6 & 0.04 & two\\
 GS9  & -17.3 & 1.20 & 20.2 & 0.9 & 0.21 & & -16.6 & 0.64 & 20.8 & 0.4 & & -16.5 & 1.00 & 22.1 & 1.3 & 0.03 & two\\
 GS10 & -18.8 & 1.27 & 20.8 & 2.5 & 0.15 & & -18.2 & 0.89 & 21.3 & 1.8 & & -18.1 & 1.00 & 23.1 & 3.9 & 0.05 & two\\
 GS11 & -20.2 & 1.05 & 20.5 & 3.1 & 0.11 & & -18.1 & 0.58 & 22.1 & 1.3 & & -20.1 & 1.00 & 20.9 & 3.3 & 0.06 & two\\
 GS12 & -19.1 & 0.94 & 20.7 & 2.4 & 0.06 & & -16.2 & 0.20 & 23.4 & 0.2 & & -19.0 & 1.00 & 20.8 & 2.5 & 0.05 & one\\
 GS13 & -18.4 & 0.87 & 21.1 & 1.6 & 0.06 & & -18.4 & 0.87 & 21.1 & 1.6 & & -8.8  & 1.00 & 30.0 & 1.2 & 0.06 & one\\
 GS14 & -19.2 & 1.61 & 18.7 & 1.7 & 0.21 & & -18.5 & 0.88 & 19.4 & 1.9 & & -18.6 & 1.00 & 21.6 & 2.8 & 0.03 & two\\
\bottomrule
\end{tabular}\\
\end{table*}

\section{Discussion}

Although an evolutionary connection between late-type and early-type dwarf galaxies was suggested over 30 yr ago
(e.g. \citealt{lin1983,kormendy1985,davies1988}), the details remain vague. Some late-type dwarf galaxies located in dense regions
may be environmentally
transformed into dEs \citep{kormendy1985,boselli2006}.
\citet{kormendy2009} and \citet{kormendy2012} showed that spheroidal galaxies are not dwarf elliptical but bulgeless S0 galaxies,
with structural parameters consistent with the irregular galaxies.
However, the difference in 
chemical abundances \citep{thuan1985,grebel2003} suggests that simple gas stripping is not enough to
transform a dIrr into a dE. In this work, we investigate the possible evolutionary connection between late- and early-type
dwarf galaxies by comparing their structural properties.

In the left-hand panel of Fig. 5, {it can be seen that}
the central surface brightness of the underlying hosts of our BCDs are fainter 
than {for} other BCD samples but consistent with the dEs and dIrrs. 
{This difference between different} BCD samples may be due to the different fitting methods
{, since we fit the two-component model to the whole galaxy rather than only an exponential function to the outer region.} 
In the right-hand panel of Fig. 5, the underlying hosts of BCDs seem to have distinctive structural properties compared to dEs and dIrrs.
Since SBPs {in the $F435W$} band only {reach} $\sim$ 26 ${\rm mag\ arcsec^{-2}}$, the structural properties may
be affected by nebular emission from {the} star formation region as discussed in \citet{Micheva2013}.  
{Meanwhile,} this distinctive distribution may also be 
due to the {luminosities used for} the underlying host of BCDs 
{and} other dwarf galaxies. 
Most {comparisons} in the literature, including Fig. 5 in this work, use the {luminosity} of a galaxy rather than the underlying host 
component. {This is not representative for BCDs, since their luminosity in the $B$ band includes a significant contribution of the central starburst region.}
If we replace the $x$-axis {for} our BCDs in Fig. 5 with {the absolute} magnitude of the outer component, more than half of the BCDs move to the region occupied
by dIrrs. Therefore, we consider
the difference of {effective radius} in {the} optical $B$ band to be unphysical.

{It is interesting to note that}
the structural parameters of the outer components of BCDs
{are in remarkable agreement with those of dEs when utilizing} the deepest {NIR imaging so far} of BCDs. 
{This consistency suggests that the 
underlying hosts of BCDs can not be distinguished from those of dEs and no significant change 
of structure is needed given the evolutionary connection between 
BCDs and dEs. In addition,
we also find an approximately common luminosity-size relationship of different types of dwarf 
galaxies which suggests a unified structural evolution for dwarf galaxies.}

In contrast to the outer {disc}, the inner {components} of BCDs are significantly different from those of dEs.
However, it should be noted that the {luminosity} of the inner {components} of BCDs and dEs is fainter than the outer component 
by $\sim$ 1.5 mag. With {fewer} data points, 
the structural parameters of {the inner component} may be more sensitive to the fitting procedure (such as initial value and parameter range)
and less robust than those {of the} outer {component}.
Nevertheless, if BCDs evolve to dEs rather than a simple cessation of 
star formation, passive fading could play a role. These likely inflict perturbations upon
the stellar body, which tend to increase the concentration,
and which could therefore lead to the higher central surface brightness, compacter size and higher S{\'e}rsic index of the inner component.

Actually, the structure of a galaxy varies at different {wavelengths} \citep{vulcani2014}. 
For example, the effective radius decreases towards longer {wavelengths}.
Therefore, the structural parameters of our BCDs derived at {the} observed {$F435W$ or $F160W$} 
band could be different {from those for the} rest-frame {$B$ or $H$} band. 
\citet{vulcani2014} investigated the wavelength-dependence of galaxy structure using a sample of bright low redshift galaxies. 
Although the dependence of 
dwarf galaxies may be different, we could roughly estimate the {redshift effect} in our comparison. 
For example, in the case of BCDs at $z$ $\sim$
0.25, the {$F435W$ and $F160W$ bands} have restframe effective {wavelengths}
of 3480 {and} 12800 \AA, respectively. 
{Although the effective radius is dependent on wavelength, the variance of wavelength is quite small with such narrow redshift range.}
In {Fig.} 13 of \citet{vulcani2014},
we find that the 
{variances} of effective radius are {both} generally small ($<$ 5 percent by eye) in the wavelength range of 3480 -- 4350 and 12800 -- 16000 \AA $\,$ 
for galaxies with blue {colour} and $n < 2.5$. The difference of S\'ersic index $n$ could be slightly larger (see {fig.} 6 in \citealt{vulcani2014}).
In a word, we consider our result about the {consistency between} the {structures} of different {types of} dwarf 
galaxies {to be} robust.

\section{Summary}
The main goal of this work is to study the structural properties of the underlying host of BCDs and 
compare {them} with other dwarf galaxies {such as} dIrrs and dEs. Unlike many previous works, which study dwarf galaxies in the nearby Universe,
we selected 34 BCDs in the CANDEL GOODS-N and GOODS-S deep fields with a median redshift of 0.2. 
With deep {NIR imaging} from {the} CANDELS survey, we obtained 
the deepest {NIR azimuthally averaged} SBP of BCDs {so far},  
{reaching} $\sim$ 26 mag ${\rm arcsec^{-2}}$ {at the 3$\sigma$ level}. 
Then we fit the SBPs with one- and two-component S\'ersic models. 
There are 16 of 33 BCDs in the $F435W$ band and 19 of 34 BCDs in the $F160W$ band which favour the two-component model. 

Comparison between the structural properties of different {types of}
dwarf galaxies offers an inspection of the possible {evolutionary} connections between them.
We derived the structural properties of the underlying {hosts}
of BCDs {in the $F435W$} band and compared them to {those} of dEs and dIrrs. 
The {effective radii} of the underlying {hosts} of BCDs {in the $B$}
band are smaller than those of
dEs and dIrrs.
This discrepancy is similar to {findings in} many previous works.
However, when {we compare} the structural properties
{in the $H$} band, the difference between BCDs and {other dwarf galaxies} 
seems to be less significant.
Furthermore, we find a remarkable agreement between the underlying {hosts} of BCDs and dEs {in the $H$} band.
{All dwarf galaxies, including dIrrs, seem to follow a similar luminosity-radius relationship, which suggests 
a unified structural evolution for dwarf galaxies.}
In contrast to the {underlying hosts}, the inner {components} of BCDs are significantly different from {those} of dEs. 
Passive fading from BCDs to dEs may be one of the possible {mechanisms} that can
{explain} the different inner structure of 
BCDs and dEs.

There are two possible reasons for the disappearance of difference in structural parameters between BCDs and other dwarf galaxies {in the $H$} band. 
\citet{Micheva2013} suggested {that}
the SBPs of BCDs {for $\mu$ in the} 24--26 mag ${\rm arcsec^{-2}}$ {range in the $B$} 
band are affected by nebular emission and therefore may not trace the underlying stellar 
population distribution. 
Besides, comparison of the {structures} of dwarf galaxies using total {luminosity}
is not {representative} for BCDs since the total {luminosities} of BCDs {in the $B$ band} include
{a} significant contribution {from} the central {star-forming} region.

With deep {NIR} photometry and detailed SBP fitting, we conclude that the 
underlying {hosts} of BCDs can not be distinguished from {those} of dEs, 
and no significant change of structure is needed given the {evolutionary} connection between 
BCDs and dEs. 

\section*{Acknowledgements}

We are grateful for the referee's insightful suggestions and comments, 
which significantly improved the quality of this paper. This work is based on observations taken by the CANDELS Multi-Cycle Treasury Program and
3D-HST Treasury Program (GO 12177 and 12328) with the NASA/ESA {\sl HST}, which is operated by the Association 
of Universities for Research in Astronomy, Inc., under NASA contract NAS5-26555. 
This work is supported by the National Natural Science Foundation of China (NSFC, nos. 11225315, 1320101002, 11433005 and 11421303), 
the Strategic Priority Research Program 'The Emergence of Cosmological Structures' of the Chinese Academy of Sciences (no. XDB09000000), 
the Specialized Research Fund for the Doctoral Program of Higher Education (SRFDP, no. 20123402110037),
and the Chinese National 973 Fundamental Science Programs (973 programme) (2015CB857004).

\end{document}